# Bicycle stabilization using mechanism optimization and digital LQR

Reza Pirayeshshirazinezhad
Texas A&M University, TX 77553

**Abstract**

This study introduces lateral pendulum as an innovative balancer design for bicycle stabilization. This pendulum, operating in the bicycle's vertical plane, enables the bicycle to remain stationary. The paper develops a dynamic model for a bicycle equipped with this lateral pendulum, using Lagrange's method, where the equations are validated with ADAMS software. The stabilization is demonstrated with traditional vertical and novel lateral pendulums, managed through a genetic-pole placement control algorithm. This approach showcases the superiority of the lateral pendulum over traditional methods, including vertical pendulums and steering the handlebar. Additionally, a Digital Linear Quadratic Regulator controller is implemented for practical application, further enhancing system stability.
***Keywords- Stability control; Genetic algorithm; LQR; Optimization, Mechanisms***

## Introduction

Research on the stabilization of bicycles has become popular in several robotic laboratories around the world. Control and modelling a bike are important since bikes have great manoeuvre, low weight, and low energy consumption. The bicycle control and dynamics literatures are extensively discussed in [1]. There has been comprehensive research in stabilizing a bicycle with forward speed. In addition, stabilizing a bicycle without speed is a much more difficult task since the speed itself is a significant stabilizing factor. However, when a bicycle stops, or its speed converges to zero, it needs to be stable in its position. In [2], a bicycle without speed is stabilized with handlebar in its upright position. In [3], the bicycle is stabilized via an inverted vertical pendulum while the bicycle is in motion. The bicycle is stabilized in a region and does not converge to zero. Also, a stop and go situation is assumed, where the bicycle is stationary and then accelerates to obtain speed. In [4], a bicycle with a handlebar and vertical pendulum that can be turned into a flywheel is stabilized when the bicycle is stationary. It is illustrated that the advantage of the flywheel is to stabilize the bicycle with large regions of stability, and the advantages of the balancer are to shift the bicycle body to avoid the obstacles and to control the bicycle to control the desired trajectory with high speed. However, one disadvantage of using a flywheel is its high energy consumption. In [5], a bicycle is stabilized with a vertical balancer, and the improvement in stability is shown when using handlebar is illustrated. In addition, path tracking is investigated when the bicycle is in motion. In [6], a lateral pendulum is used to stabilize a bicycle for path tracking. However, there are several simplifications in the dynamic equations and the bicycle has forward speed. These oversimplifications reduce the capability of optimizing the stabilizing mechanism and designing an optimal control solution.

Generally, a vertical inverted pendulum or handlebar is used to stabilize the bicycle. As it is mentioned in [2], only a small region roll angle of bicycle can be stabilized with handlebar. Similarly, this is the same issue with the vertical balancer since we have constraints on our DC motor for its speed and torque capacities. In this research, a lateral pendulum, which is better in stabilizing bicycles, is implemented on a bicycle. To stabilize a bicycle with a vertical pendulum, when a bicycle is falling, assuming the angular velocity and angle of the pendulum would be zero, the bicycle inevitably needs to use the reaction torque and reaction acceleration of the pendulum to return to the upright position. These reaction torque and acceleration make the pendulum fall, which in turn leads to the instability of the bicycle. In addition, the angular velocity of the pendulum makes the bicycle unstable. A controller should make a balance between these reactions and stabilize the bicycle. To tackle this problem and maximize the stabilizable set, a lateral pendulum is proposed since the reaction torque only twist the bicycle in its horizontal plane and is removed by reaction forces on tires. However, the reaction angular acceleration by pendulum imposed on the bicycle makes the bicycle. To solve the problem, pendulum should be placed near the ground. By placing the pendulum near the ground, the effect of angular velocity and acceleration decreases and so the effect of pendulum mass increases. If the pendulum is placed on the top of the bicycle, the need for higher torques increases which decreases the advantage of using a lateral pendulum over the vertical pendulum.

The contribution of this paper is to design an optimal mechanism and optimal control systems for bicycle stabilization.

In the following, first, the full dynamic equations of bicycle with a later pendulum are driven and validated. Next, both vertical pendulum and lateral pendulum mechanism and controllers are optimized. The advantage of using a lateral pendulum to stabilize the bicycle over vertical pendulum and the handlebar is shown. Finally, an optimal digital controller, considering the DC motor voltage as the input to the system, is implemented on the bicycle.

## Equations

In this section, we seek to drive the equations of a bicycle with a lateral pendulum rotating in the horizontal plane of the bicycle. To drive the equations, the Lagrange method is implemented, and the equations are verified with ADAMS software. For the ADAMS model, it is assumed that bicycle is rotating around X axis of the B coordinate system, which is attached to the bottom of the rear wheel, and the lateral pendulum is rotating around z axis of the E coordinate system, which is attached to the lateral pendulum, and handlebar rotation is neglected. In the derived equations, the variation of the bicycle angle around X axis of the B coordinate system is declared by $x_r$ and the lateral pendulum angle variation around z axis of the E coordinate system is defined by n. In the following, how equations are derived with Lagrange method is discussed and the equations results for an especial case are compared with the ADAMS bicycle model outputs for verification.

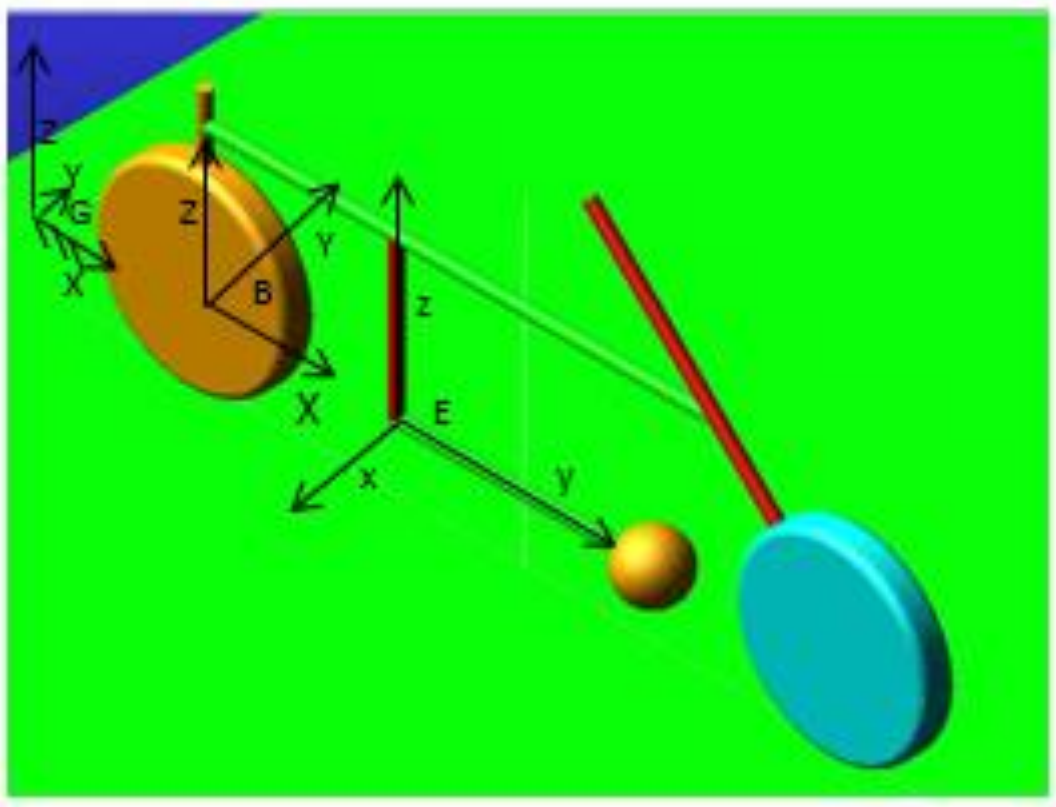

(a)

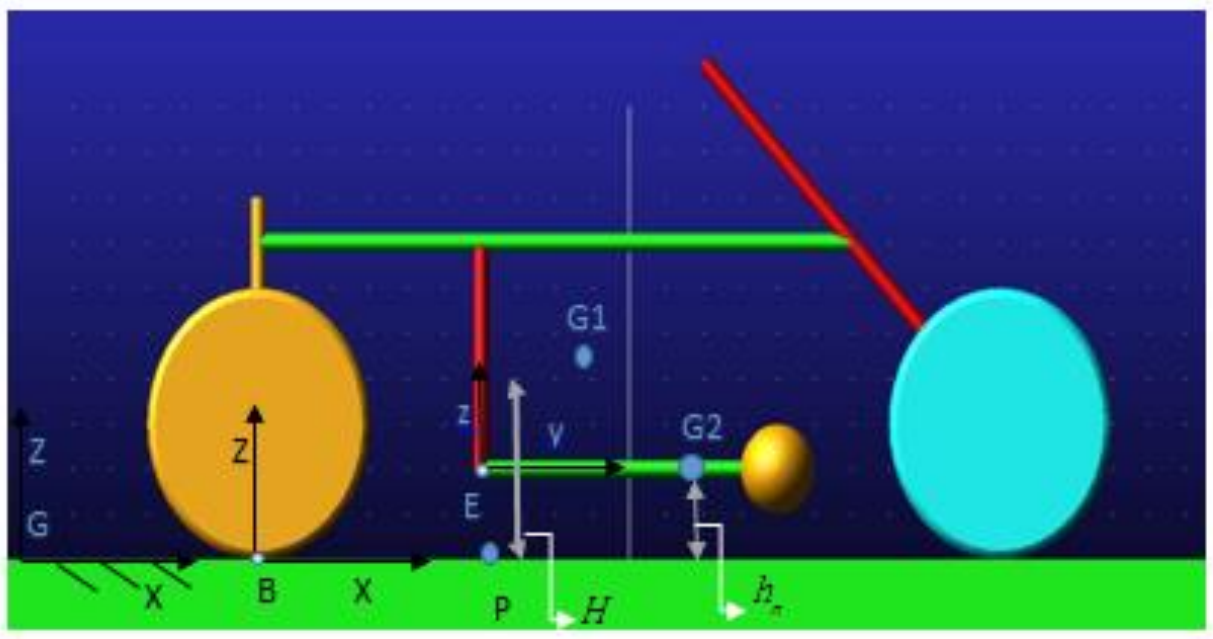

(b)

Fig.1 (a) ADAMS model-side view and (b) model-front view

### Lagrange dynamical equations

In Lagrange method, the two coordinates (n and $x_r$ ) are generalized coordinates and potential and kinetic energy are based on these coordinates. The following parameters are utilized in the equations:

| Parameter | Parameter description |
|---|---|
| $I_G\,(kg-m^2)$ | Inertial momentum of the lateral pendulum around the center of mass |
| $I_r\,(kg-m^2)$ | Inertial momentum of bicycle around the x axis of B coordinate system |
| $m_n\,(kg)$ | lateral pendulum mass |
| $h_n\,(m)$ | Height of the lateral pendulum from ground |
| $L\,(m)$ | Length of the lateral pendulum center of mass from hinge of the pendulum |
| $M\,(kg)$ | Bicycle weigh |
| $H\,(m)$ | Height of the center of mass of the bicycle from ground |
| $g(\frac{m}{s^2})$ | gravity |
| U(N-m) | Torque provided by DC motor |
| $\overline{MH}$ | $\overline{MH} = MH + m_n h_n$ |
| m(kg) | Vertical pendulum weigh |
| h(m) | Vertical pendulum height |

$$KE = 0.5 \times (m_n \vec{V}_{G2}.\vec{V}_{G2} + \overline{w}^T I_G \overline{w} + I_r \dot{x}_r^2) \qquad (1)$$

$$\overline{w} = \dot{x}_r \sin n\vec{i} + \dot{x}_r \cos n\vec{j} + \dot{n}\vec{k}$$
$$\vec{V}_{G2} = \vec{V}_E + \overline{w} \times \vec{r}_{G2/E} \qquad (2)$$

$$I_G = \begin{bmatrix} \overline{I}_{xx} & 0 & 0 \\ 0 & \overline{I}_{yy} & 0 \\ 0 & 0 & \overline{I}_{zz} \end{bmatrix} \qquad (3)$$

In the given equations, $\overline{w}$ is the angular velocity of the lateral pendulum and $\vec{V}_{G2}$ is the velocity of the center of mass of the lateral pendulum.
By some substitutions, the following kinetic energy is obtained:

$$KE = \frac{1}{2} m_n h_n^2 \dot{x}_r^2 (\cos n)^2 + \frac{1}{2} m_n L^2 \dot{n}^2$$
$$-m_n L h_n \dot{x}_r \dot{n} cosn + \frac{1}{2} m_n h_n^2 \dot{x}_r^2 (\sin n)^2 \qquad (4)$$
$$+\frac{1}{2} m_n L^2 \dot{x}_r^2 (\sin n)^2 + \frac{1}{2} \overline{I}_{xx} \dot{x}_r^2 \sin n^2$$
$$+\frac{1}{2} \overline{I}_{yy} \dot{x}_r^2 \cos n^2 + \frac{1}{2} \overline{I}_{zz} \dot{n}^2$$

And potential energy as

$$PE = \overline{MH} cosx_r - m_n Lgsin(n) \sin x_r \qquad (5)$$

Lagrangian function is defined as $L = KE - PE$ . After substituting (1) and (2) into the Lagrangian function, the following Lagrangian function is derived.

$$L = 0.5\overline{\overline{I}}_r \dot{x}_r^2 + 0.5\overline{\overline{I}}_z \dot{n}^2 - m_n L h_n \dot{x}_r \dot{n} \cos n +$$
$$0.5\overline{\overline{I}}_{xx} \dot{x}_r^2 (\sin n)^2 + 0.5\overline{I}_{yy} \dot{x}_r^2 (\cos n)^2 \qquad (6)$$
$$-\overline{MH} g \cos x_r - m_n Lg \sin n \sin x_r$$

Next, the equations of motion is obtained from

$$\frac{d}{dt}\left(\frac{\partial L}{\partial \dot{q}_i}\right) - \frac{\partial L}{\partial q_i} = Q_i \qquad (7)$$

Where $q_i$ is generalized coordinate and $Q_i$ is the corresponding generalized force. Generalized force of the system is obtained as the following:

$$\partial w = Q_{x_r} \partial x_r + Q_n \partial n$$
$$Q_{x_r} = 0 \qquad (8)$$
$$Q_n = U$$

Generalized force of $(x_r)$ coordinate is zero, and the generalized force of n is the applied torque exerted on the pendulum by the DC motor. The reason is that the only active torque that violates the n generalized coordinate is U, while there is no active torque and force violating the $(x_r)$ coordinate. Substituting the equations (6) and (8) in (8) leads to the following equations of motion:

$$\ddot{x}_r = [\dot{x}_r \dot{n}_r sin 2n \left(\overline{I}_{yy} - \overline{\overline{I}}_{xx}\right) +$$
$$m_n L h_n \ddot{n} cosn - m_n L h_n \dot{n}^2 \sin(n)$$
$$-m_n Lgsin(n) \cos x_r + \overline{MH} gsinx_r]/ \qquad (9)$$
$$[\overline{\overline{I}}_{xx} \sin n^2 + \overline{I}_{yy} \cos n^2 + \overline{\overline{I}}_r]$$

$$\ddot{n} = [U + m_n L h_n \ddot{x}_r cosn +$$
$$0.5\dot{x}_r^2 \sin n^2 \left(-\overline{I}_{yy} + \overline{\overline{I}}_{xx}\right) - \qquad (10)$$
$$m_n Lgcos(n) \sin x_r]/\overline{\overline{I}}_{zz}$$

Where the parameters are defined as

$$\overline{\overline{I}}_r = I_r + m_n h_n^2 \qquad (11)$$
$$\overline{\overline{I}}_{zz} = \overline{I}_{zz} + m_n L^2$$
$$\overline{\overline{I}}_{xx} = \overline{I}_{xx} + m_n L^2$$

To verify the equations, an ADAMS model with arbitrary physical parameters and initial conditions is developed. Next, the equations are, with the same parameters and initial conditions, simulated in the Simulink of MATLAB. The ADAMS model output is illustrated with red line and MATLAB Simulink results with blue line. This is done for different initial conditions and torques, and the results are the same, which proves the accuracy of the equations. In the following plots, one of the different comparisons, the input torque is assumed to be $0.1N/m$ and $\dot{x}_r = 8\deg/s$ .

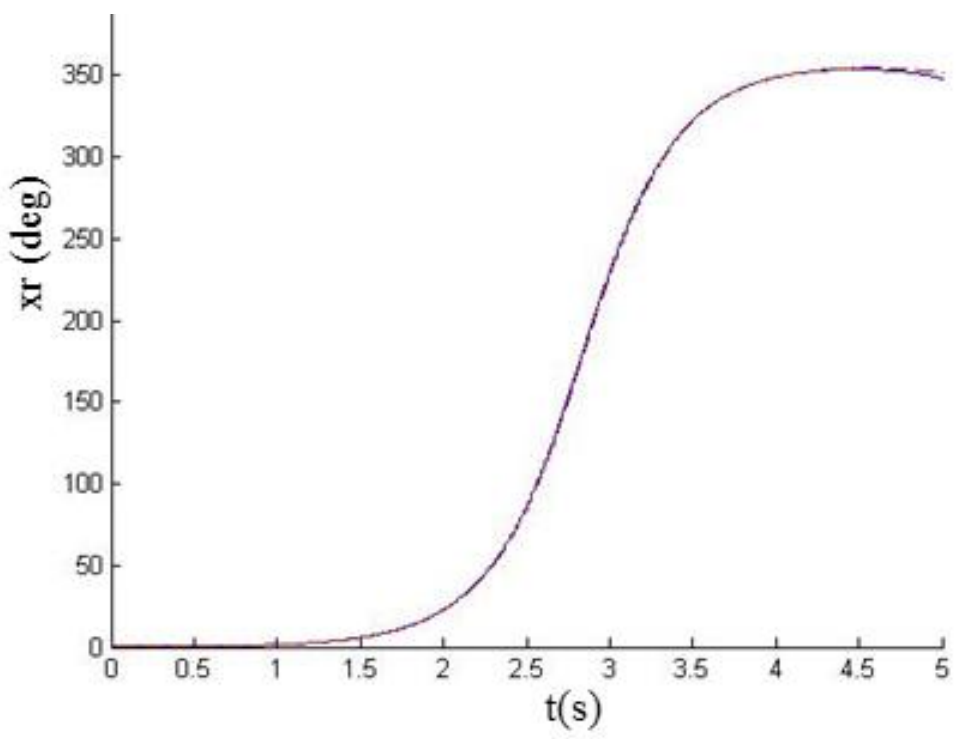

Fig. 1 - Roll angle

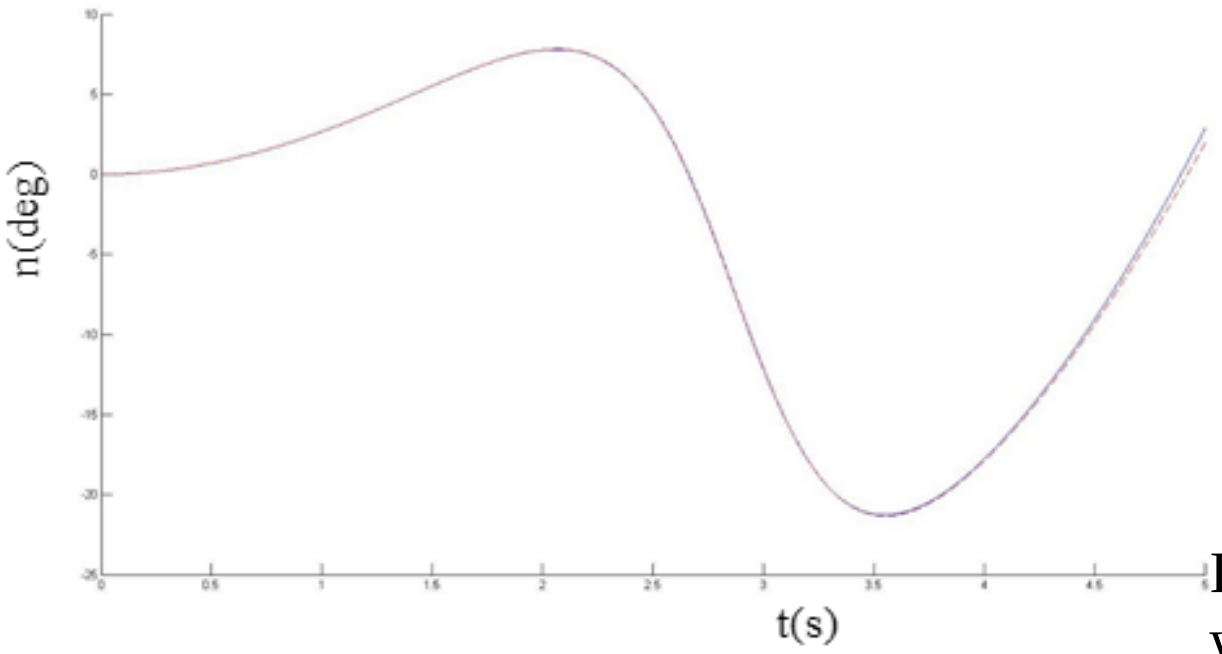

Fig. 2 - Pendulum angle

The figures clearly illustrate the accuracy of the equations.
The chosen bicycle for the next simulations is as the following.

$$\bar{\bar{I}}_{zz} = \bar{I}_{zz} + m_n L^2$$
$$\bar{\bar{I}}_{xx} = \bar{I}_{xx} + m_n L^2$$
$$\bar{I}_{yy} = 0.0055\,kg - m^2$$
$$\bar{I}_{xx} = 0.0295\,kg - m^2$$
$$\bar{I}_{zz} = 0.0295\,kg - m^2$$
$$h_n = 13cm$$
$$L = 15cm$$
$$I_r = 0.9kg - m^2$$
$$\bar{\bar{I}}_r = I_r + m_n h_n^2$$
$$H = 0.2667\,cm$$
$$M = 9$$
$$I_G = \begin{bmatrix} \bar{I}_{xx} & 0 & 0 \\ 0 & \bar{I}_{yy} & 0 \\ 0 & 0 & \bar{I}_{zz} \end{bmatrix}$$

*Bicycle with vertical pendulum equations*

In this section, the equations of motion of a bicycle with vertical pendulum is taken from [3] and [4]. A stationary handlebar for the bicycle is assumed. In the following, the equations are given.

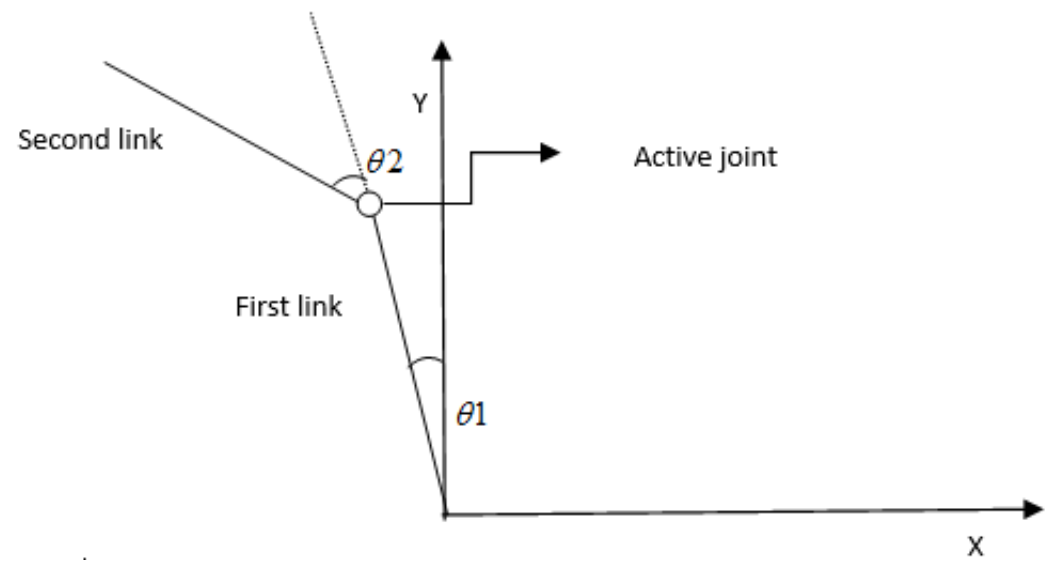

Fig. 3 – bicycle with pendulum schematic

$$M(q)\ddot{q}+C(q,\dot{q})+G(q)=U \quad (11)$$

$$M(q)=\begin{bmatrix} a+g+2b\cos(\theta 2) & a+b\cos(\theta 2) \\ a+b\cos(\theta 2) & g \end{bmatrix}$$

$$C(q,\dot{q})=\begin{bmatrix} -b\dot{\theta}2(2\dot{\theta}1+\dot{\theta}2)\sin(\theta 2) \\ b\dot{\theta}1^2\sin(\theta 2) \end{bmatrix} \quad (12)$$

$$G(q)=-\begin{bmatrix} k1\sin(\theta 1)+k2\sin(\theta 1+\theta 2) \\ k2\sin(\theta 1+\theta 2) \end{bmatrix}$$

$$U=[0,U2]^T$$

$$a=m_1L_{g1}{}^2+I1+m_2L_1{}^2$$

$$b=m_2L_1L_{g2}$$

$$g=m_2L_{g2}{}^2+I2$$

$$k1=(m_1L_{g1}+m_2L_1)g_o$$

$$k2=m2L_{g2}g_o$$

In the next section, the pendulum physical parameters are optimized by genetic algorithm, and the controller is obtained by genetic-pole placement.

## Controlling algorithm

In this paper, we seek to stabilize the bicycle by hybrid genetic-pole placement. DC motor's characteristics are its torque and the maximum power it can provide. The higher a DC motor can provide torque and maximum power, the more expensive and heavier the DC motor is. In order to find the best pendulum physical parameters and poles, first the mass, length and poles are chosen as the input of the genetic algorithm and the target function is the area under the curve of the bicycle angle and pendulum angular velocity. In fact, different target functions are utilized during the optimization to find the best physical parameters and poles to minimize the target function, which in turn leads to defining cheaper and lighter DC motors, low overshoot, and less settling time for the bicycle to stabilize.

For the maximum torque of the DC motor, saturation is considered since a limitation for the torque of the DC motor would suffice when the angular velocity is minimized during optimization. The reasons are as follows:

1) We seek to increase the region of stability for the bicycle, which leads to higher torques and pendulum angular velocities. Accordingly, by considering saturation for DC motor and minimizing the pendulum angular velocity, we obtain the approximate optimized DC motor characteristics.
2) Furthermore, the more sophisticated the target function is, the longer optimization takes.

After finding the topology of the pendulum, the pendulum physical parameters are tuned according to the limitations of the system and better performance. The reason we change and tune them is because of two reasons:

1) If the initial condition for the system is assumed high, Genetic cannot find proper poles to stabilize the system perfectly. Consequently, first, Genetic finds the physical parameters and then, they are changed to the desired values.
2) It becomes more feasible to define the topology of the pendulum by trial and error when genetics have defined the pendulum physical parameters first.

In the following, first, the bicycle with lateral pendulum is stabilized. Next, the bicycle is stabilized by a vertical pendulum and handlebar. Finally, the results are compared and the advantage of using a lateral pendulum is discussed.

## Bicycle with lateral pendulum

In this section, the bicycle with lateral pendulum is stabilized with Genetic- pole placemen algorithm. Genetic defines the poles as the following:

poles=[-19.7726,-15.4665,-21.7665,-11.1548]

And the pendulum physical parameters with some modifications as:

$m_n=6.6kg$

$h_n=13cm$

$L=15cm$

In the following figures, the bicycle is stabilized with the initial angular velocity $\dot{x}_r=30\deg/\sec$ and maximum torque 6N-m.

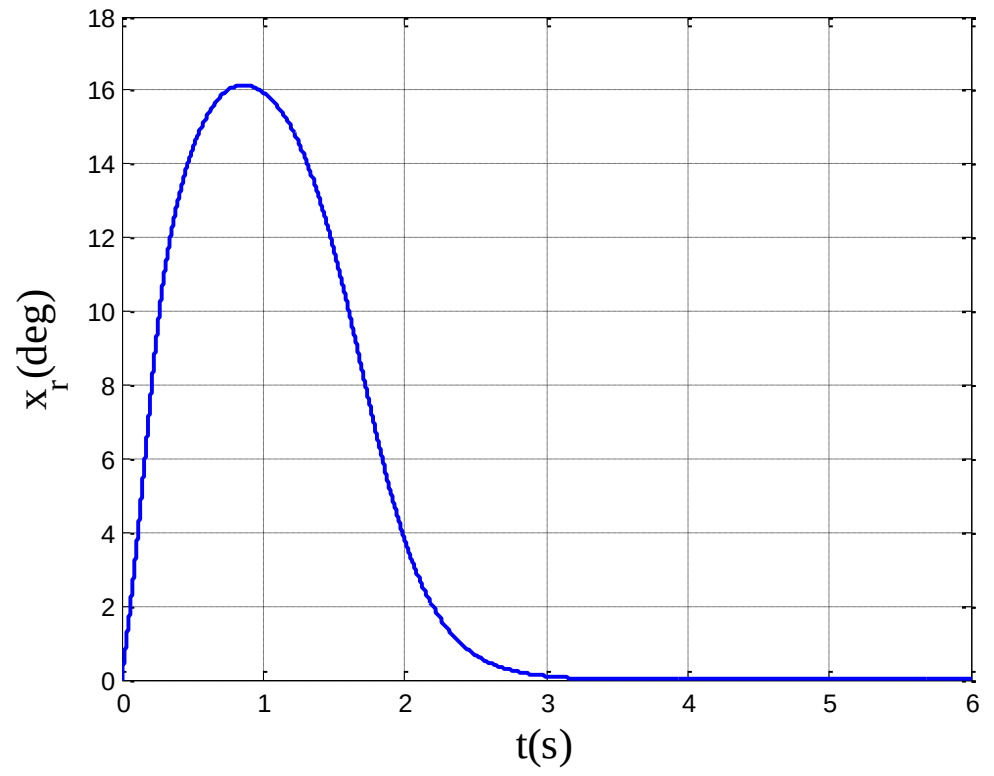

Fig. 4- Roll angle

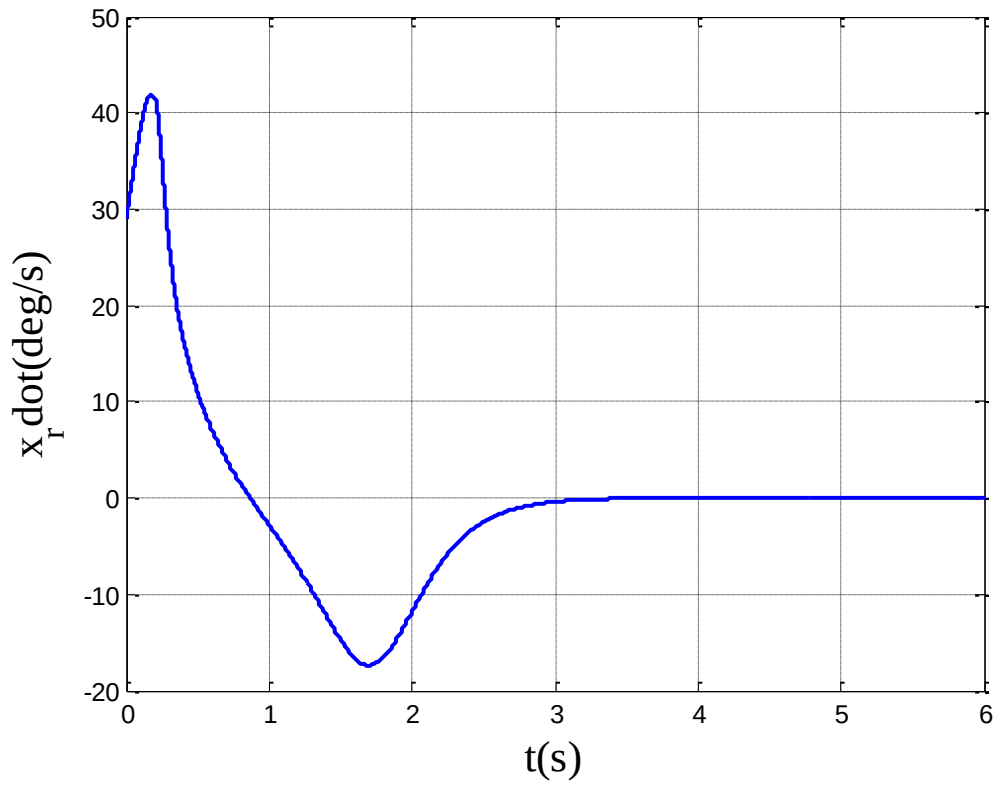

Fig. 5- Roll angular velocity

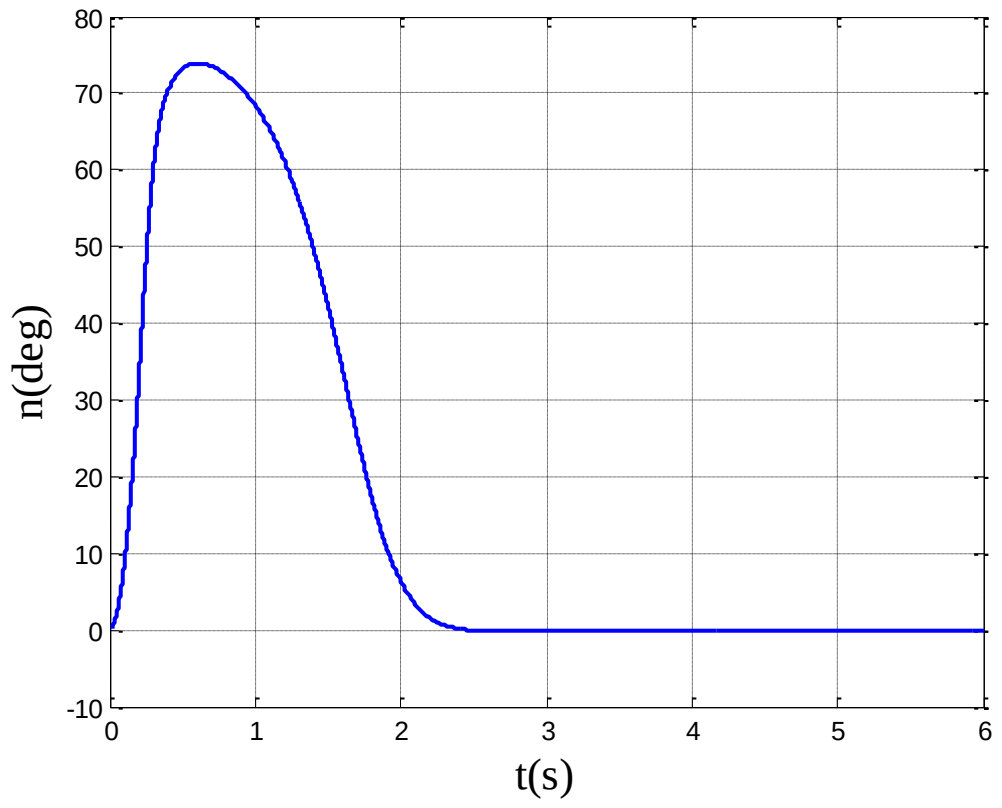

Fig. 6- Pendulum angle

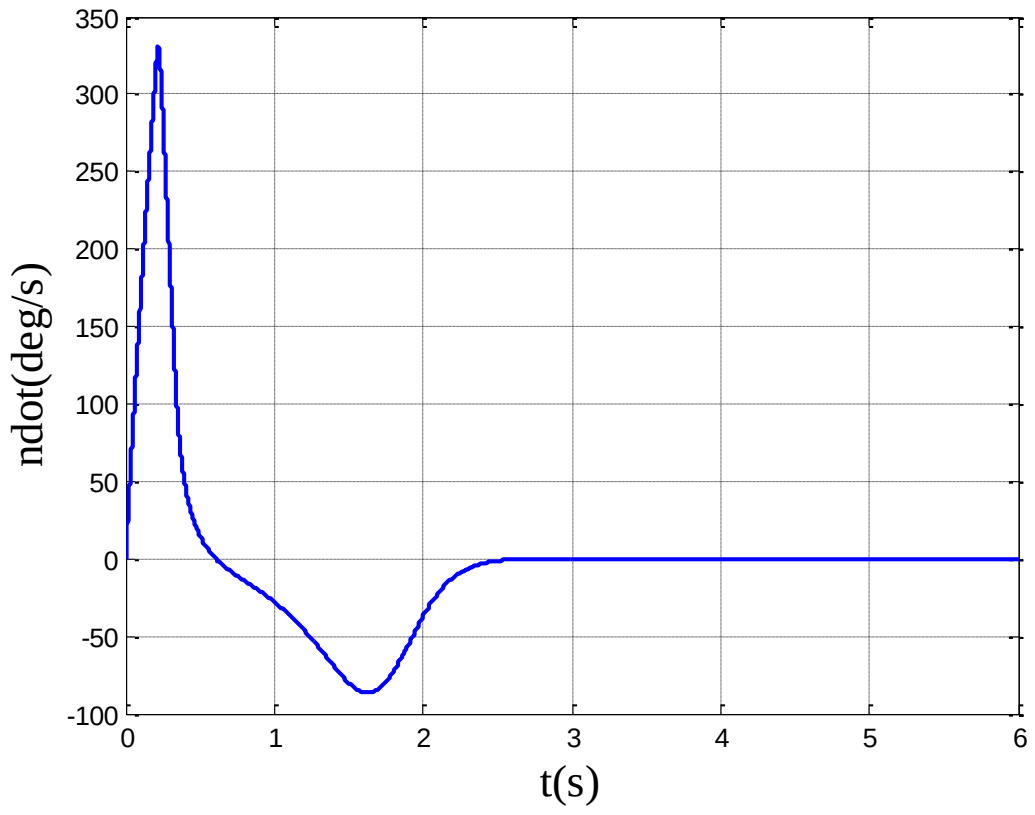

Fig. 7- Pendulum angular velocity

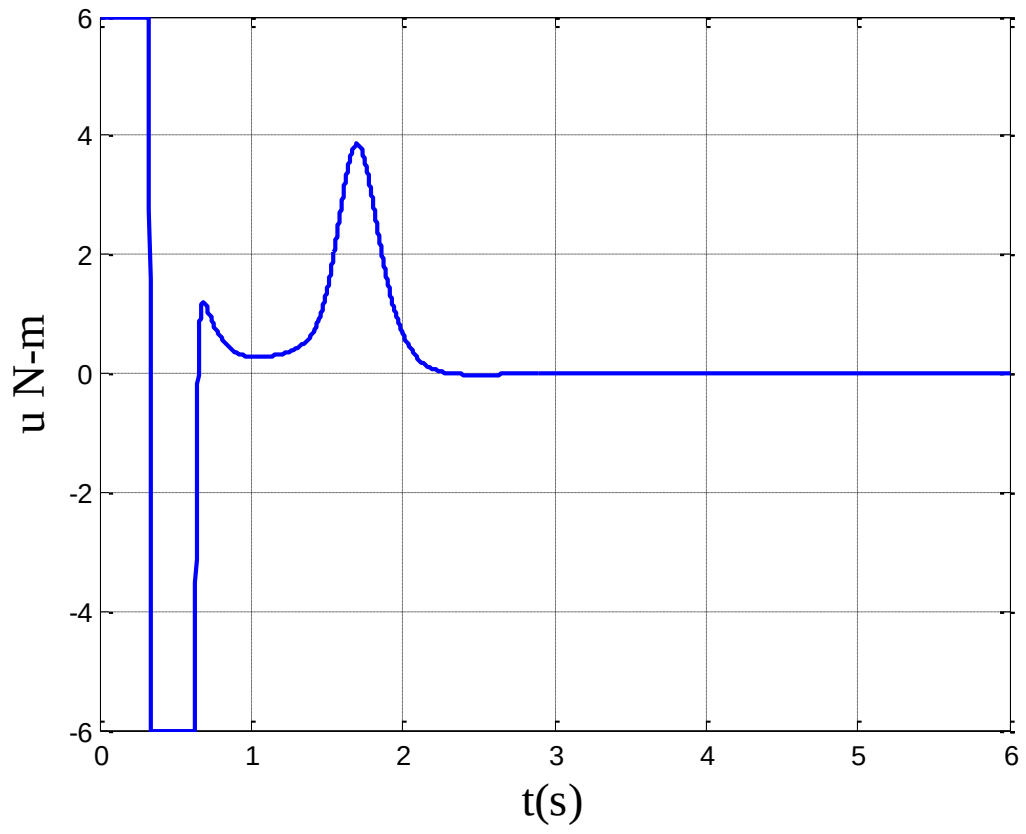

Fig. 8- Torque

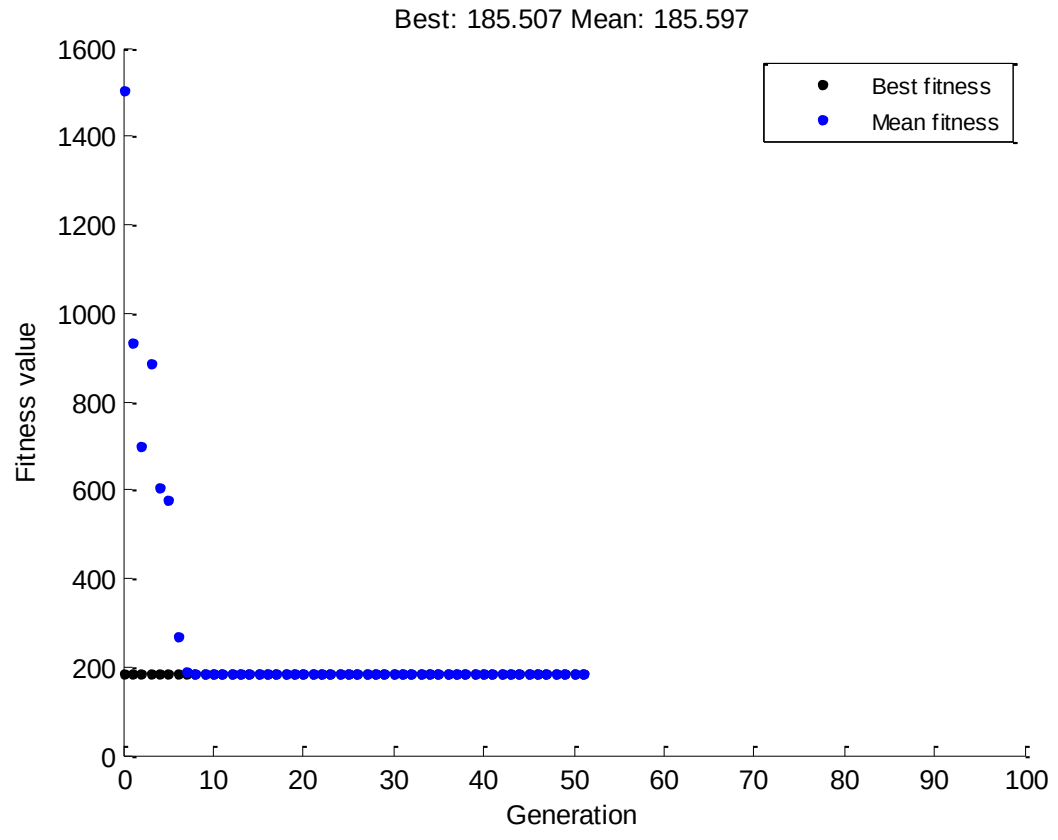

Fig. 9- fitness

The results clearly show the efficiency of the controlling algorithm and the capacity of the lateral pendulum to stabilize the bicycle.

## Bicycle with vertical pendulum

In this section, the bicycle with vertical pendulum is stabilized with Genetic- pole placemen algorithm. Genetic defines the poles as the following:

poles=[-19.7726,-15.4665,-21.7665,-11.1548]

And the pendulum physical parameters with some modification as:

$m = 0.3739 kg$

$h = 0.3354 m$

The physical parameters are feasible for implementation on a real bicycle.

In the following figures, the bicycle is stabilized with the initial angular velocity $\dot{x}_r = 3 \deg/\sec$ and maximum torque 6N-m.

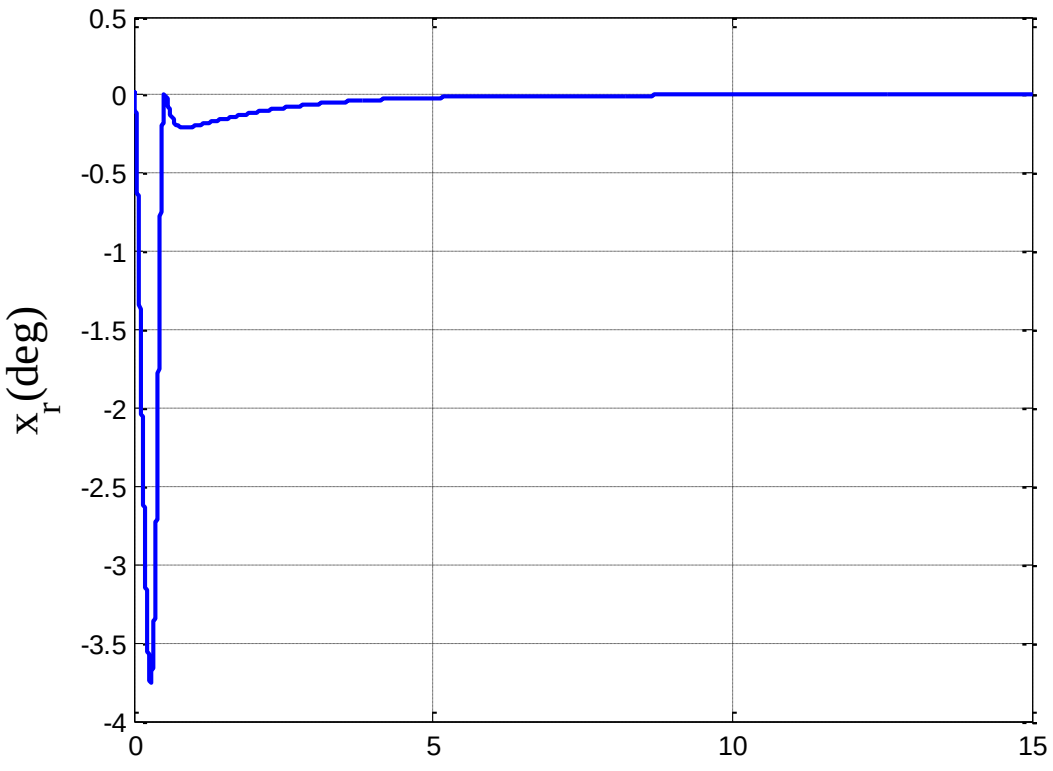

Fig. 11- Roll angle

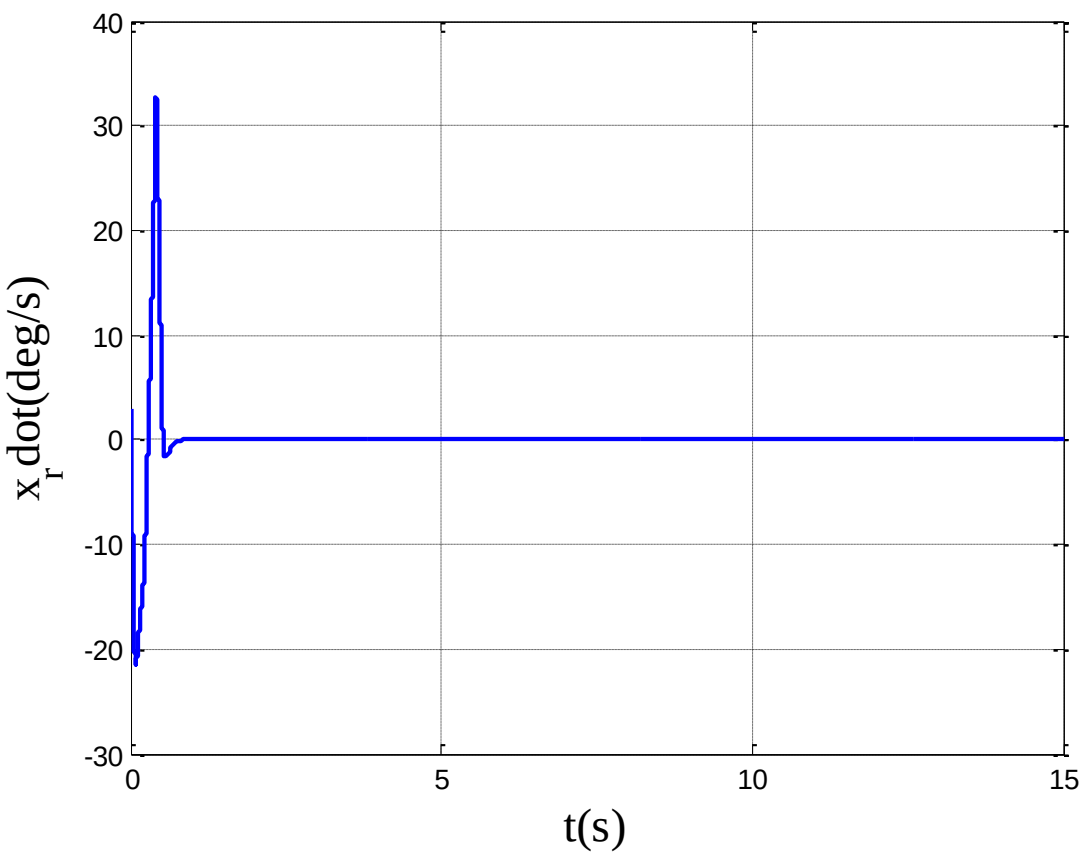

Fig.12- Roll angular velocity

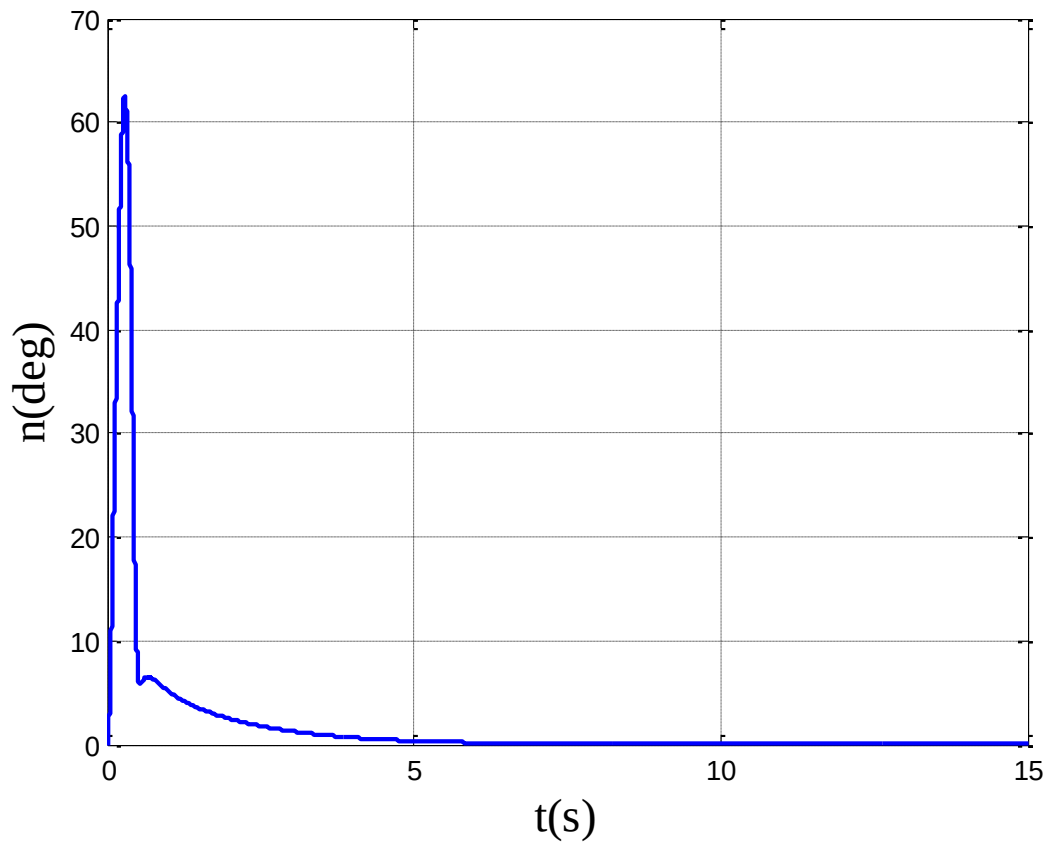

Fig. 13- Pendulum angle

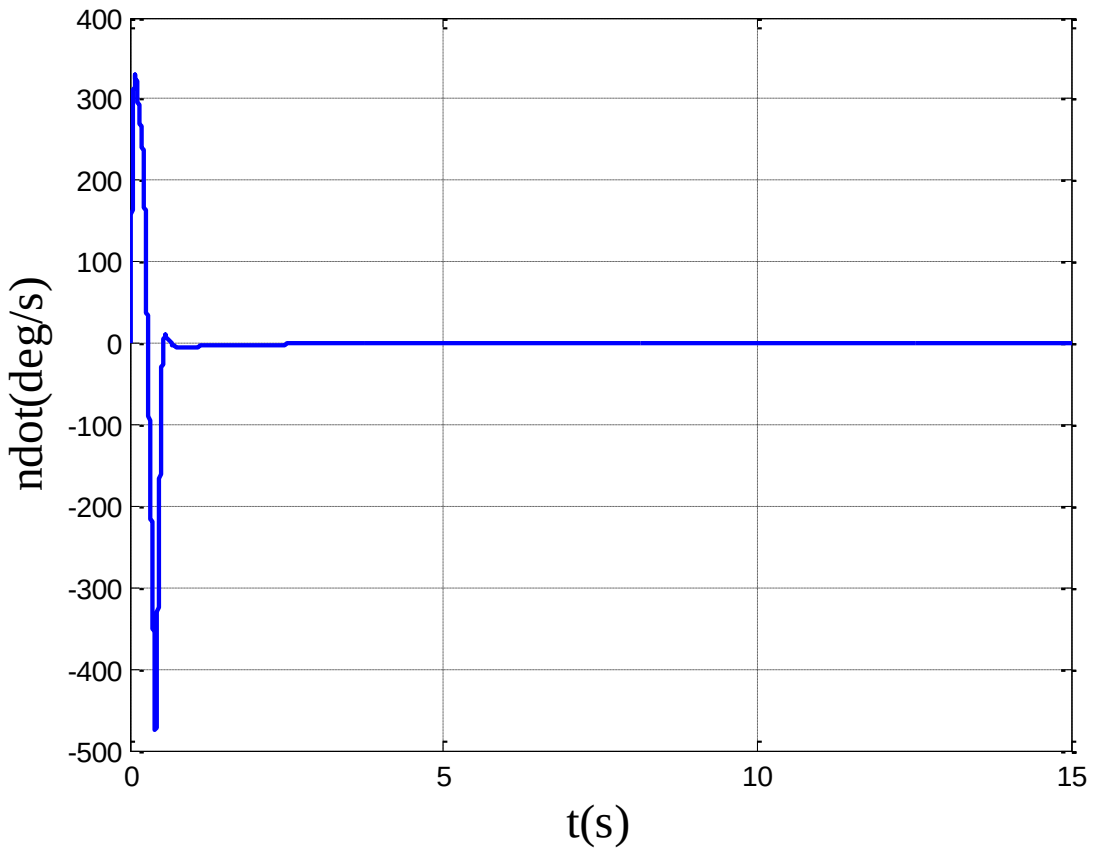

Fig. 14- Pendulum angular velocity

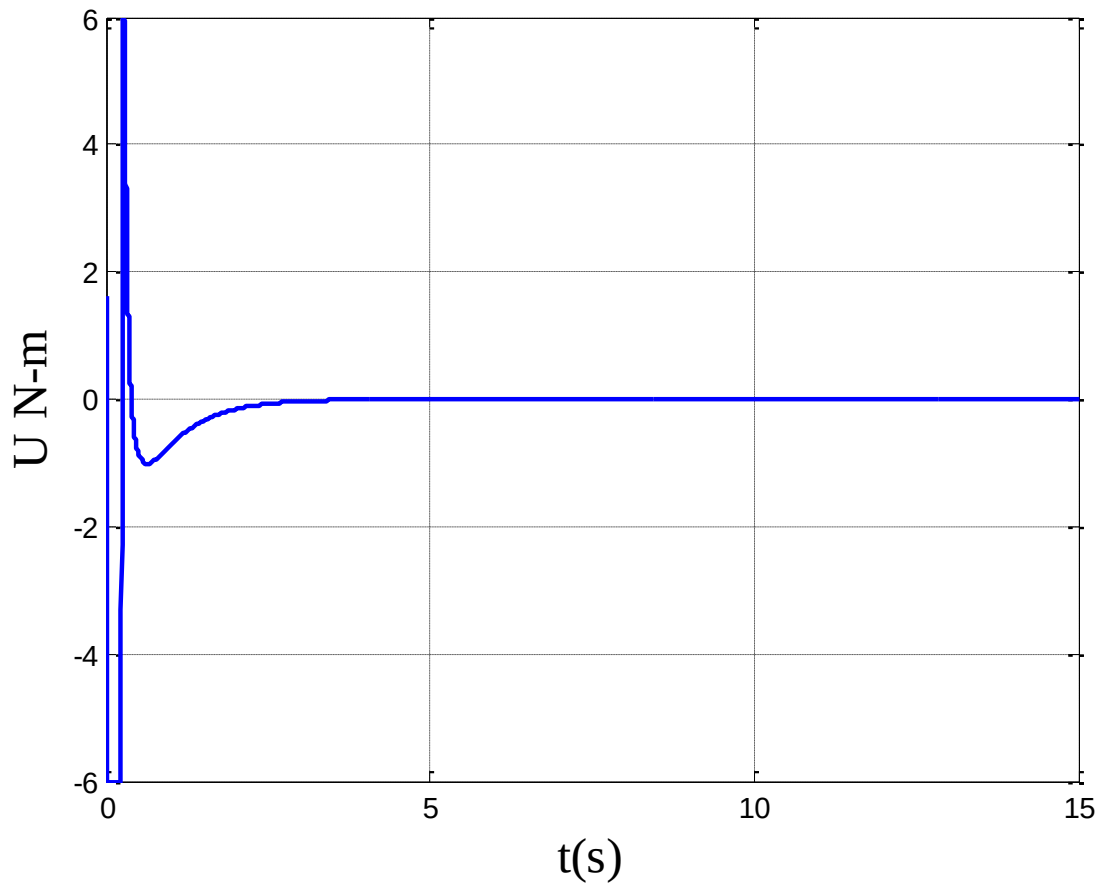

Fig. 105-Torque

And the fitness of the genetic algorithm converges as follows:

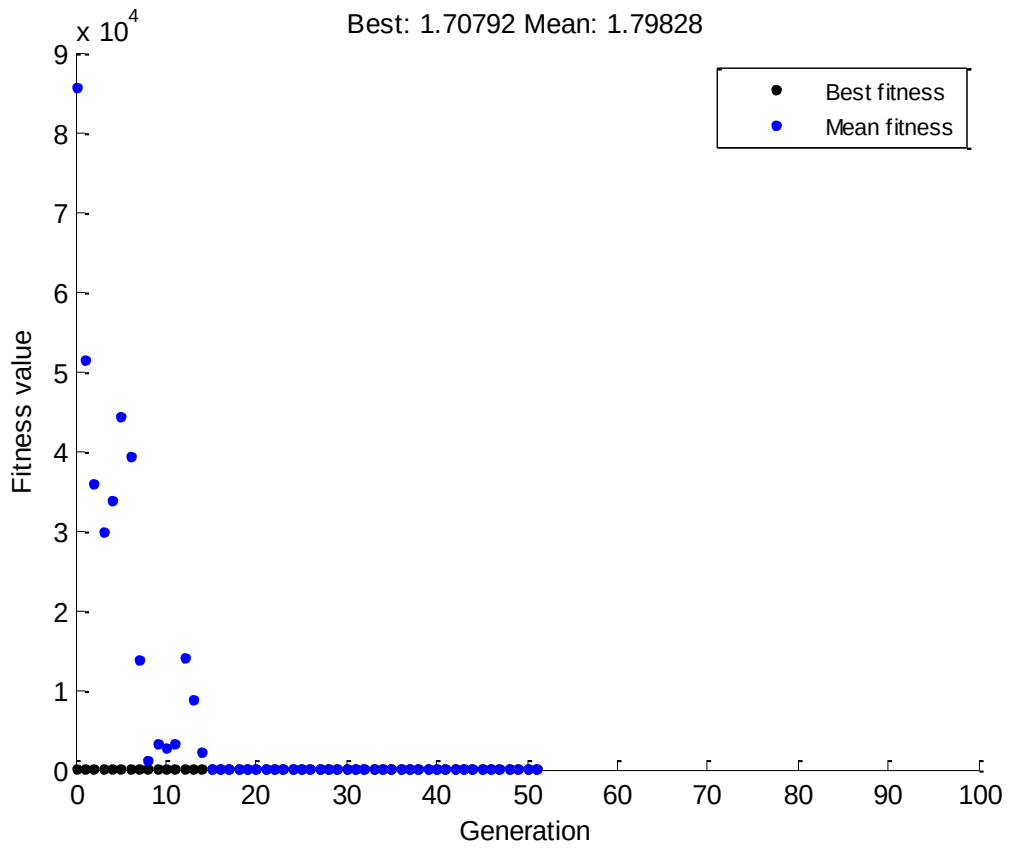

Fig. 16- fitness

In the given figures, the bicycle was stabilized with the small initial angular velocity. However, if the angular is increased up to 15deg/sec, the bicycle cannot be stabilized and the bicycle roll angle changes according to the following:

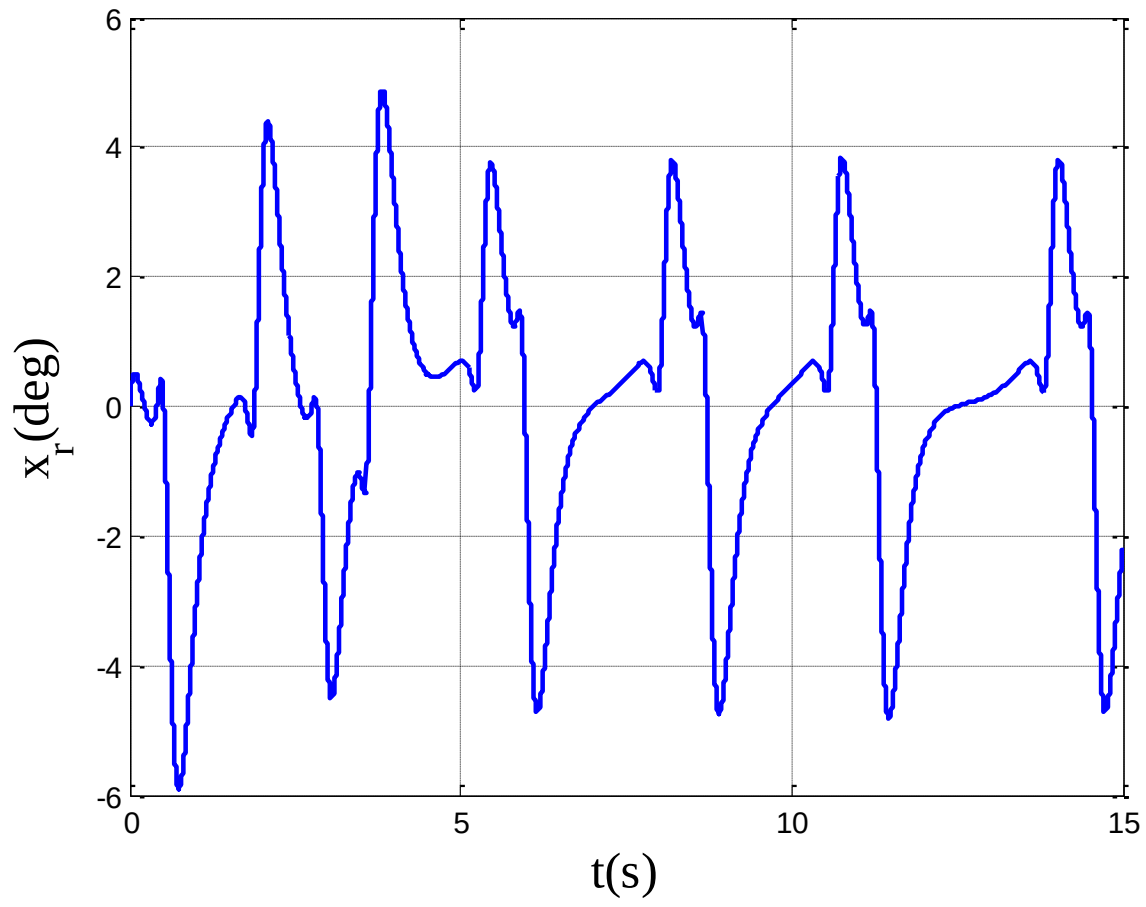

Fig. 17- Roll angle

The performance of the system is not satisfactory, and the bicycle is unstable. In this case, other target functions are also utilized but the bicycle could not become stable.

Lateral pendulum can stabilize the bicycle with bicycle initial angular velocity 30deg/sec and maximum pendulum angular velocity around 330deg/sec, while the vertical pendulum can only stabilize the bicycle with bicycle initial angular velocity around 3deg/sec and maximum pendulum angular velocity around 450deg/sec, and it cannot stabilize the bicycle with the bicycle initial angular velocity around 15deg/sec. This clearly proves the advantage of using lateral pendulum over vertical pendulum.

## Comparing the stability of using handlebar and lateral pendulum

In this section, the performance of using handlebar to stabilize the bicycle is compared with lateral pendulum to do so. To compare them, the bicycle in [2] is stabilized with lateral pendulum with genetic-pole placement and the advantage of using lateral pendulum is discussed.

The physical parameters are:

$\bar{I} = 0.0551 kg - m^2$

$I_r = 4.2786 kg - m^2$

$m_n = 17 kg$

$h_n = 14 cm$

$L = 20 cm$

$H = 0.27 cm$

$M = 34.9 kg$

$x_r = 6\deg$ (*initial condition*)

$Poles = [-22.5121 \quad -17.5332 \quad -15.6349 \quad -12.1521]$

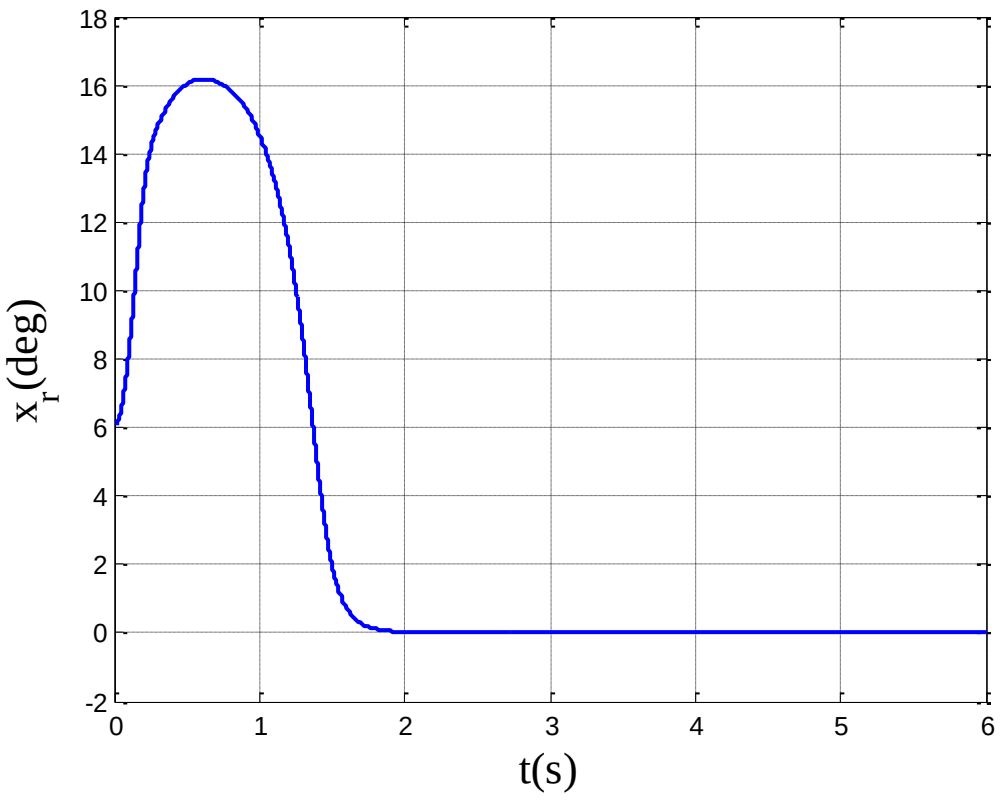

Fig. 18- Roll angle

and the given diagram for the roll angle of the stationary bicycle given in [2] is as follows.

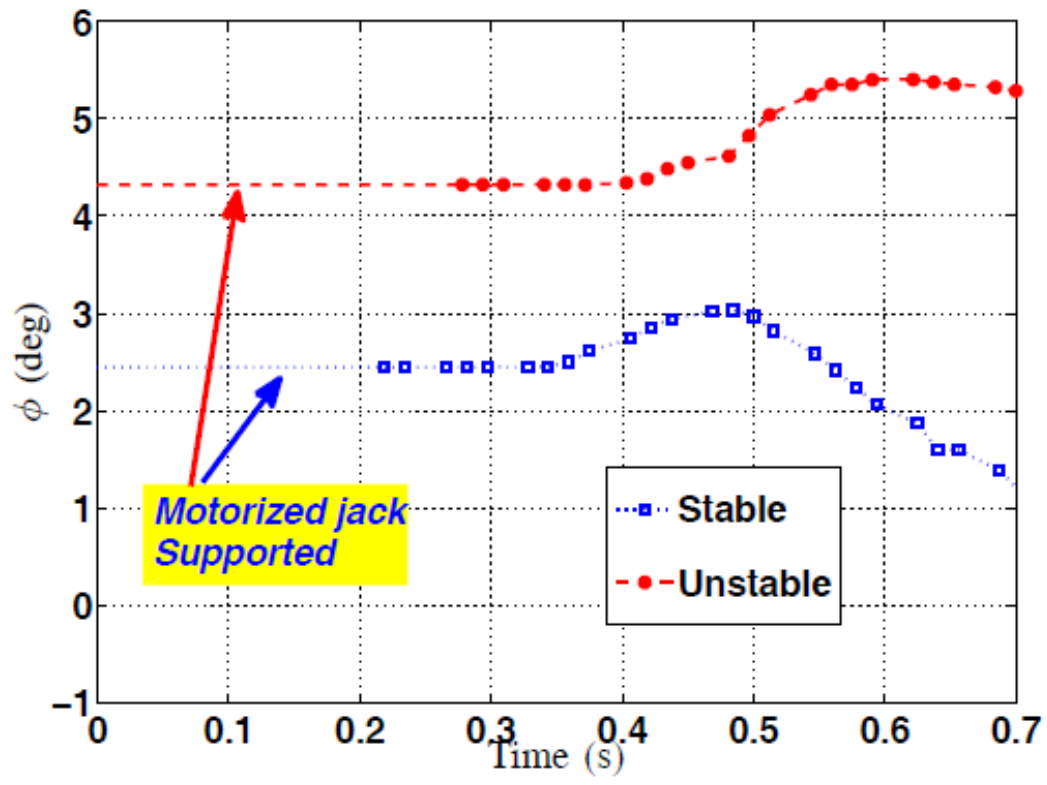

Fig. 19- Roll angle

As it is clear, Bicycle with lateral pendulum had better performance since handlebar could only stabilize the bicycle with bicycle initial roll angle $x_r = 2.5\deg$, while lateral pendulum could stabilize bicycle with bicycle initial angle $x_r = 6\deg$. This clearly shows the advantage of using lateral pendulum over handlebar to stabilize bicycle when it is stationary.

**Implementation**

The experimental implementation of the proposed lateral pendulum on the bicycle is shown in Fig. 20.

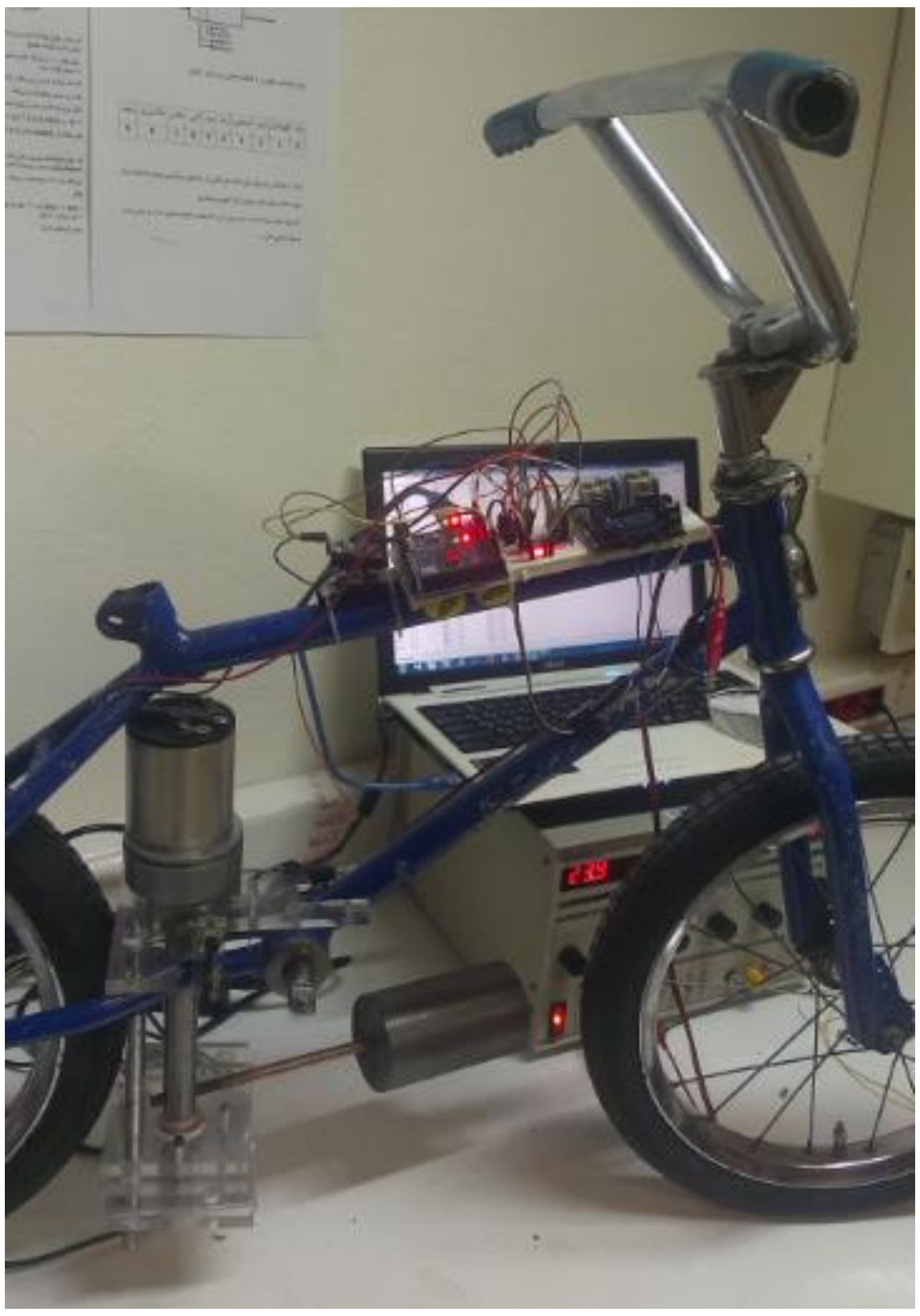

Fig. 20- Experimental setup

In this part, we aim to consider the voltage of the DC motor as the input of the system, since in real implementation the input to the DC motor is the voltage and we have control over it. Accordingly, the DC motor model is considered in the state equations. The DC motor properties are obtained by experiment as the following:

R=14.5657;

L=0.9650;

$k_t$=5.5;

And the DC motor equations is:

$$\dot{U} = -R/LU + K_t/LVin - k_t^2/L\dot{n}$$

The state space is:

$$\begin{bmatrix} \dot{x}_r \\ \ddot{x}_r \\ \dot{n} \\ \ddot{n} \\ \dot{U} \end{bmatrix} = f(x) \begin{bmatrix} x_r \\ \dot{x}_r \\ n \\ \dot{n} \\ U \end{bmatrix} + BVin$$

*Controlling algorithm*

In this part, we seek to implement an optimal liner controller on the system. Accordingly, first the system is linearized by Jacobian function in MATLAB and discretized by C2dm function with the 0.01sec discretization sampling time and ZOH method.

The discrete linearized state space is as the following:

$$x(k+1) = Ax(k) + BVin(k)$$

$$y(k) = Cx(k) + DVin(k)$$

$$A = \begin{bmatrix} 1.0000 & 0.0010 & -0.0000 & -0.0000 & 0.0001 \\ 0.0223 & 1.0011 & -0.0517 & -0.0037 & 0.0211 \\ -0.0051 & -0.0000 & 1.0000 & 0.0098 & 0.0018 \\ -1.0151 & -0.0051 & -0.0101 & 0.9994 & 0.3495 \\ 0.1528 & 0.0005 & 0.0149 & -0.2851 & 0.8060 \end{bmatrix}$$

$$B = \begin{bmatrix} 0.0000 \\ 0.0006 \\ 0.0000 \\ 0.0103 \\ 0.0518 \end{bmatrix}$$

$$C = \begin{bmatrix} 1 & 0 & 0 & 0 & 0 \\ 0 & 1 & 0 & 0 & 0 \\ 0 & 0 & 1 & 0 & 0 \\ 0 & 0 & 0 & 1 & 0 \\ 0 & 0 & 0 & 0 & 1 \end{bmatrix}$$

$$D = \begin{bmatrix} 0 \\ 0 \\ 0 \\ 0 \\ 0 \end{bmatrix}$$

Second, the optimal controller is chosen to be Digital Linear Quadratic Regulator. This controller has the quadratic cost function defined as the following:

$$J(u) = \sum_{n=1}^{\infty} (x[n]^T Qx[n] + u[n]^T Ru[n] + 2x[n]^T Nu[n])$$

which is minimized by the controller input as the following:

$$U = -R^{-1}B'\bar{P}x(t)$$

Where $\overline{p}$ is found from solving Riccati equation as the following:

$$0 = A'\overline{P} + \overline{P}A + Q - \overline{P}BR^{-1}B'\overline{P}$$

Since we seek to have low maximum torque, low maximum DC motor angular velocity, and low energy consumption, a second target function is defined to be minimized as the following:

y=0.01*trapz(abs(xr))+30*max(abs(ndot))+

65*max(abs(torque))+0.1*trapz(abs(ndot)*abs(torque))

The DC motor power is obtained by

$$Power = T(torqueN-m)*w(rad/\sec)$$

The algorithm minimizing the target function is Genetic-Pattern search. The weights are considered according to the importance of each element that needs to be minimized. Take, for example, the maximum amount of torque and maximum angular velocity. They are more important than for example the lateral angle of the bicycle since the less maximum torque and the angular velocity of the DC motor is, the lighter and cheaper the DC motor is.

Since the maximum voltage of the considered DC motor is 30 volts, the saturation considered for the input is 30 volts.

The fitness function for this optimization in Fig. 21.

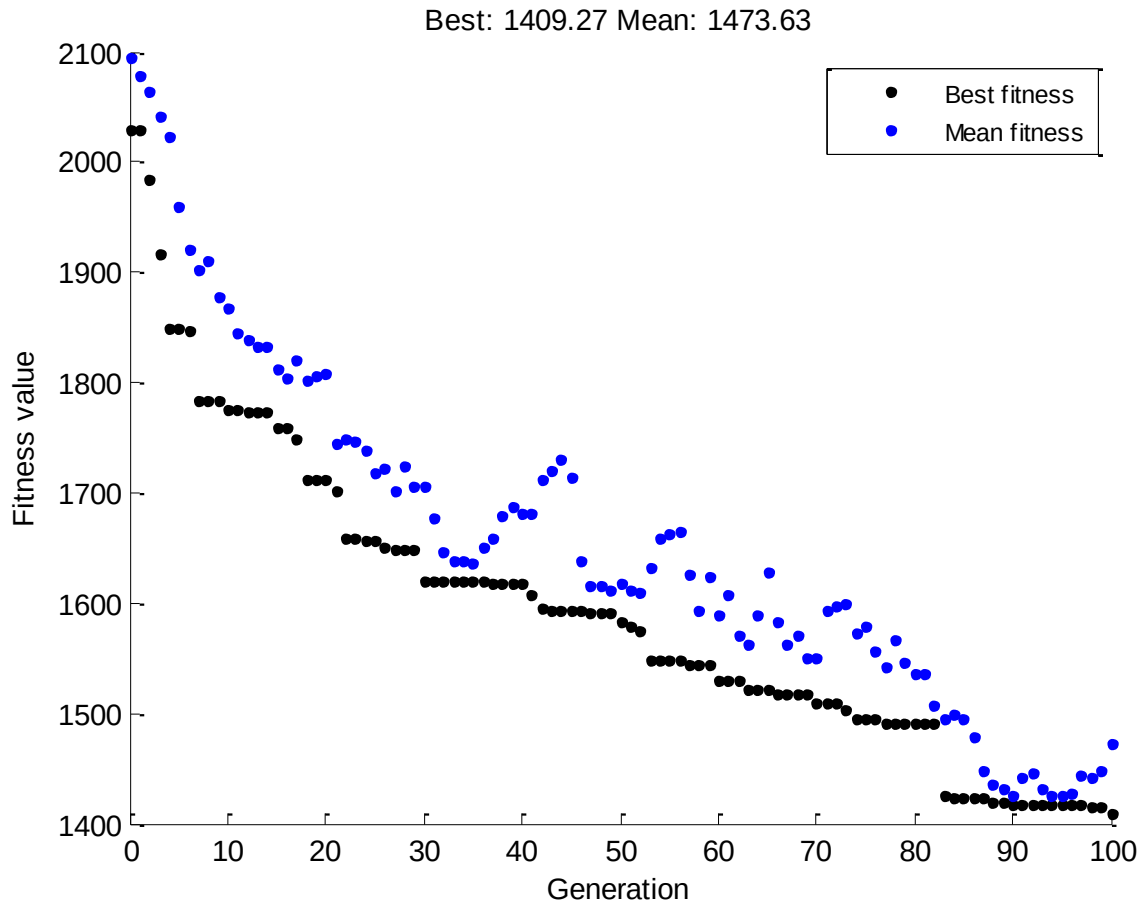

Fig. 21- Fitness function for the optimization

And this optimization leads to the following Q and R:

$$Q = \begin{bmatrix} a1 & 0 & 0 & 0 & 0 \\ 0 & a2 & 0 & 0 & 0 \\ 0 & 0 & a3 & 0 & 0 \\ 0 & 0 & 0 & a4 & 0 \\ 0 & 0 & 0 & 0 & a5 \end{bmatrix}$$

$$R = a6$$

$$a1 = 5.9619e+06$$

$$a2 = 2.4446$$

$$a3 = 3.4464$$

$$a4 = 7.8203e+05$$

$$a5 = 9.7913e+07$$

$$a6 = 2.2404$$

In Fig. 22, the bicycle is stabilized with the initial angular velocity $\dot{x}_r = 20\deg/\sec$ and maximum torque 6N-m.

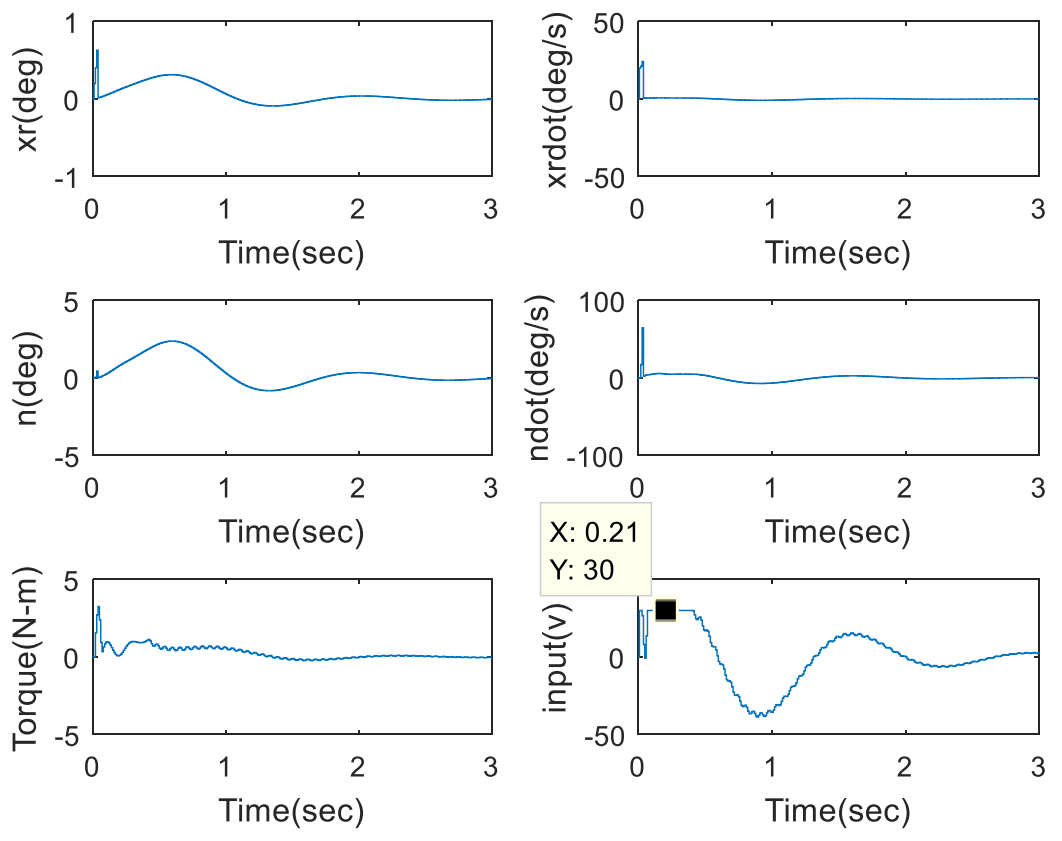

Fig. 22- States convergence

As it is clear from the figures, the system is stabilized in 3 seconds and overshoot of the bicycle is less than 1 degree.

## Conclusion

In this paper, first, a new actuator for stabilizing a stationary bicycle is proposed and how this new mechanism is advantageous to other common mechanisms is discussed. The lateral pendulum for stabilizing bicycle is better since with a good design, bicycle can be stabilized with DC motors with less torques and power. Next, we stabilized a bicycle with a genetic-pole placement with both vertical and lateral pendulum. Afterwards, the advantage of using a lateral pendulum was illustrated versus vertical pendulum and also handlebar to stabilize the bicycle. Finally, the bicycle is stabilized be Digital Linear Quadratic Regulator controller and shows satisfactory attitude.

## Acknowledgments

This paper expands upon research initially conducted during my master’s studies at the university of Tehran. I am deeply thankful for the foundational knowledge and resources provided by the institution, which played a crucial role in the genesis of this work.

I extend my heartfelt gratitude to my master's thesis advisor, Dr. Masoud Shariat Panahi, whose guidance, insights, and unwavering support have been invaluable in shaping both this research and my academic trajectory.